# Erbium silicide growth in the presence of residual oxygen

Nicolas Reckinger,[a,z] Xiaohui Tang,[a] Sylvie Godey,[b] Emmanuel Dubois,[b] Adam Łaszcz,[c] Jacek Ratajczak,[c] Alexandru Vlad,[a] Constantin Augustin Duțu,[a] and Jean-Pierre Raskin[a]

[a]*ICTEAM institute, Université catholique de Louvain (UCL), Place du Levant 3, 1348 Louvain-la-Neuve, Belgium*

[b]*IEMN/ISEN, UMR CNRS 8520, Avenue Poincaré, Cité Scientifique, 59652 Villeneuve d'Ascq Cedex, France*

[c]*Institute of Electron Technology, Aleja Lotników 32/46, 02-668, Warsaw, Poland*

[z]E-mail: nicolas.reckinger@uclouvain.be

The chemical changes of Ti/Er/*n*-Si(100) stacks evaporated in high vacuum and grown *ex situ* by rapid thermal annealing were scrutinized. The emphasis was laid on the evolution with the annealing temperature of (i) the Er-Si solid-state reaction and (ii) the penetration of oxygen into Ti and its subsequent interaction with Er. For that sake, three categories of specimens were analyzed: as-deposited, annealed at 300 °C, and annealed at 600 °C. It was found that the presence of residual oxygen into the annealing atmosphere resulted in a substantial oxidation of the Er film surface, irrespective of the annealing temperature. However, the part of the Er film in intimate contact with the Si bulk formed a silicide (amorphous at 300 °C and crystalline at 600 °C) invariably free of oxygen, as testified by x-ray photoelectron spectroscopy depth profiling and Schottky barrier height extraction of 0.3 eV at 600 °C. This proves that, even if Er is highly sensitive to oxygen contamination, the formation of low SBH Er silicide contacts on *n*-Si is quite robust. Finally, the production of stripped oxygen-free Er silicide was demonstrated after process optimization.

In the recent years, the continuous trend to device miniaturization continuously requires the introduction of additional elements from Mendeleev's table in complementary metal-oxide-semiconductor (CMOS) technology, for instance, rare-earth (RE) metals. RE oxides[1,2,3,4] and silicates[5] find application as gate dielectrics because of their high relative permittivity. On the other hand, RE silicides are materials of interest for the formation of ohmic or low Schottky barrier contacts over lowly doped *n*-type Si substrates. They are indeed known to present the lowest Schottky barrier heights (SBH) to electrons, amounting to 0.27-0.28 eV.[6,7,8] In that context, the RE silicides are particularly attractive for source/drain contacts in Schottky barrier metal-oxide-semiconductor field-effect transistors (SBMOSFET)[9,10,11] and as materials for infrared detectors.[12]

RE metals are highly prone to oxidation because of their high affinity for oxygen,[13] useful property for the formation of oxides and silicates. However, the formation of one of these compounds is often accompanied with the formation of the other(s). For instance, RE oxides are prone to transform into silicides or silicates after the unavoidable high-temperature annealing in standard CMOS technology, a source of degradation in the dielectrics. Conversely, RE silicates or oxides are readily formed during RE silicide growth, which can be detrimental to the silicide resistivity and to its SBH on *n*-Si since a very small oxygen content at the interface with Si causes a substantial SBH rise.[14] A first possibility to avoid oxidation is to perform the Er-Si solid-state reaction in ultrahigh vacuum (UHV) conditions.[6,15,16] That method is very suitable to fundamental studies but less appropriate to industrial applications. To relax the constraints on the fabrication of RE silicides, several groups proposed to cover REs with a capping layer (one-shot deposition without breaking vacuum) and to grow the silicide by *ex situ* annealing.[8,17,18,19]

In this article, we propose to scrutinize the chemical changes of thin erbium (Er) films deposited under high vacuum conditions on Si in the presence of residual oxygen after *ex situ* thermal treatment by x-ray photoelectron spectroscopy (XPS). The objective is first to investigate the penetration of oxygen into the stack and to determine which reaction predominates at the interface with Si. It is also to assess the capability to grow low SBH Er silicide/*n*-Si contacts in nonideal process conditions (oxygen-contaminated annealing ambience and high vacuum deposition). Solutions to process oxygen-free Er silicide are also examined. In addition, the joint formation of Er oxide, silicide, and silicate expected in such a stack is exploited to conduct a thorough XPS analysis of these compounds, thereby enhancing the scarce published information. The obtained data are also critically compared with a relevant selection of the results disseminated in the literature, intended as a database that could be useful to scientists interested in the formation of Er silicide or Er oxides/silicates.

## Experimental

The starting wafers are lowly doped *n*-Si(100) bulk wafers (phosphorus-doped with a resistivity of 5-10 Ω cm). The Si wafers are cleaned in two successive baths of sulfuric peroxide mixture (SPM, $H_2SO_4$:$H_2O_2$ in ratio 5:2) for 10 min each, rinsed in de-ionized water for 10 min and dried by centrifugation under $N_2$. Then, the wafers are dipped into 2% hydrofluoric acid to remove Si oxide grown during the SPM cleaning, rinsed, and dried as previously mentioned.

The samples are next introduced into the evaporator immediately after the preparation steps. The metal deposition is performed in an e-beam evaporator operating under high vacuum (HV) with a base pressure of $\sim10^{-7}$ mbar. In order to clean the Er and Ti targets, 50 nm of metal are evaporated just before deposition (with the shutter closed). Ti is chosen on purpose as capping material since it is permeable to oxygen. 25 nm of Er and different Ti thicknesses (10, 15, and 50 nm) are then successively deposited. Sample 1 (with 10 nm of Ti) is left as-deposited. Five samples (2 to 6) with various Ti thicknesses ($t_{Ti}$) are heated *ex situ* by rapid thermal annealing (RTA) in $N_2$ under diverse annealing conditions, immediately after extraction from the evaporator: (i) sample 2 with $t_{Ti}$ = 10 nm annealed at 300 °C for 2 min with a 2 min initial purge, (ii) sample 3 with $t_{Ti}$ = 15 nm annealed at 600 °C for 2 min with a 2 min initial purge, (iii) sample 4 with $t_{Ti}$ = 50 nm annealed at 600 °C for 2 min with a 2 min initial purge, (iv) sample 5 with $t_{Ti}$ = 10 nm annealed at 600 °C for 2 min with a 10 min initial purge, and (v) sample 6 with $t_{Ti}$ = 50 nm annealed at 600 °C for 2 min with a 10 min initial purge. To help the reader, Table I recapitulates the stack thicknesses and the RTA conditions.

The chemical modifications within the samples are carefully studied by means of *ex situ* XPS. Transmission electron microscopy (TEM), x-ray diffraction (XRD) and electrical characterization are used to complement and to consolidate the XPS results. The XPS analyses are performed with a Physical Electronics 5600 spectrometer fitted in an UHV chamber (base pressure $\sim10^{-10}$ mbar). We use a monochromatized Al x-ray source (hν = 1486.6 eV) and the detection angle is 45° with respect to the sample surface normal. Depth profile analysis is realized by sputtering using an $Ar^+$ ion gun operated at 1 keV with a beam raster of 5×5 $mm^2$. The pass energy is set to 47 eV for depth profiles and 23 eV for more resolved spectra, leading to overall resolutions as measured from the full width at half-maximum of Ag $3d_{5/2}$ about 0.9 eV and 0.7 eV, respectively.

## Results and discussion

Core level spectra are recorded from erbium (Er 4$d$), oxygen (O 1$s$), silicon (Si 2$p$ or Si 2$s$), and titanium (Ti 2$p$). Intensity depth profiles are determined from the corresponding spectra. Capital letters with distinct superscripts for each sample are used to mark characteristic positions along the intensity profile. The choice of the markers is made arbitrarily but always corresponds to the appearance or to the vanishing of a compound or a chemical species. The information that we can extract from intensity profiles is relatively poor in the specific case of Er. Quantitative analysis from such profiles is particularly delicate because of the relative difficulty to obtain accurate intensities, since Er possesses a complex spectrum for which the determination of the integration boundaries is ambiguous. Moreover, discrepancies can arise from different sputter yields and burial effects. To push the investigation further, we rather focus on the binding energies (BE) changes, mainly from Er 4$d$ and Si 2$p$/2$s$ spectra. We also follow the modifications of the spectrum shape since it brings valuable additional information, especially for Er. BEs relative to metallic, silicided or oxidized (i.e. ErO$_x$ with 0 < x ≤ 1.5) Er can be found in[20,21,22,23,24,25,26,27,28,29,30] To make the interpretation of the recorded spectra easier, Table II summarizes the Er 4$d_{5/2}$, O 1$s$, and Si 2$p$/2$s$ BEs found in the above-mentioned papers. In addition, Table II also contains typical peak positions relevant to Si-O and Ti-O compounds[31,32,33,34,35,36,37,38] that could presumably form during RTA. When a BE is given in the text below, the reader is systematically invited to refer to Table II for more information about the corresponding reference. In addition, all BEs recorded at the different markers are gathered in Table III, to facilitate the reading of the article.[1]

**A. As-deposited Ti/Er/Si(100) stack (sample 1)**

In this work, we deliberately do not consider the silicidation of uncapped Er layers by RTA because Er silicide films produced in these conditions present poor electrical and morphological properties.[8,19,39] In addition, we also perform an XPS depth profiling of a thin Er film (20 nm) inserted immediately after evaporation into the XPS analysis chamber to minimize the exposure to ambient air. It is observed that, even with that precaution, the Er film is oxidized in depth.

Figure 1 displays the XPS intensity depth profile for sample 1 (with markers A to G). It can be clearly observed that oxygen can penetrate into the Ti layer but remains essentially concentrated at the surface since the oxygen amount is seen to quickly decrease. The top surface (marker A) is characterized by a Ti 2$p_{3/2}$ BE of 458.9 eV. Elemental Ti features a Ti 2$p_{3/2}$ core level peak centered at 454 eV while its position is reported to shift to 458.9-459 eV when Ti dioxide (TiO$_2$) is considered. The presence of TiO$_2$ is therefore unambiguously identified at the top surface of the stack, expectable due to the strong reactivity of Ti in the presence of oxygen. Oxygen most probably originates from ambient air, by diffusion in the Ti cap during the transfer between the evaporation and the introduction in the XPS analysis chamber.

---

[1]For Ti and Er, the BEs considered in the text are those of the Ti 2$p_{3/2}$ and Er 4$d_{5/2}$ peaks respectively.

Deep inside the top layer (marker B), near the interface with the Er film, Ti is weakly or not affected by oxidation, as confirmed by the Ti 2$p_{3/2}$ peak position which is centered at 453.7 eV. Simultaneously, a small amount of Er is detected. It is interesting here to check for the efficiency of the Ti capping layer to preserve Er from oxidation after a short period of time in ambient air. Before examining the chemical state of deposited Er at label B, we provide BE reference values for elemental and oxidized Er obtained from a sixth sample with a thick Er layer, deposited and analyzed in the same conditions. Figure 2 displays the corresponding normalized Er 4$d$ core level spectra after background subtraction. For elemental Er (deep inside the film), we extract an Er 4$d_{5/2}$ BE value of 167.1 eV, in fair agreement with the values found by Swami *et al.* (167.4 eV) and Kennou *et al.* (167.6 eV). In addition, Swami *et al.* mentioned that the Er 4$d_{5/2}$ peak of elemental Er presents a well-resolved satellite peak[2] (with a core level peak of 169 eV), contrary to oxidized Er, as also observed here at 169.2 eV in Fig. 2. From the analysis of the oxidized surface of the same Er thick film, Er 4$d_{5/2}$ and O 1$s$ core levels are found to be centered at 169.3 and 531.2 eV, respectively. Previously published material gives corresponding BEs comprised between 168.6 and 170.4 eV for Er 4$d_{5/2}$ and in the 530.6-531.6 eV range for O 1$s$ (see Table II). Even though those data are rather scattered, the Er 4$d_{5/2}$ peak of oxidized Er can be readily recognized since it presents two plainly identifiable characteristics comparatively to metallic Er (see Fig. 2): (1) both main and satellite peaks merge in the oxidized form and (2) the energy of the main peak is systematically about 2 eV higher than its metallic counterpart. The inspection of the Er 4$d$ spectrum at marker B gives a BE equal to 167.7 eV. A likely interpretation is that Er at marker B is a mixture of oxidized and mainly elemental Er.

Slightly deeper, starting from marker C, the BE of the Er 4$d_{5/2}$ main peak becomes constant and is determined to be 167.1 eV, as exhibited in Fig. 3a presenting the evolution of the Er 4$d_{5/2}$ and Si 2$p$ BEs versus the sputter time. Figure 3b, which gives the Er 4$d$ normalized spectrum at some selected positions after background subtraction, illustrates also the appearance of the doublet peak structure from marker C, thereby confirming the metallic state of Er. But it is however worth noting that the match is not perfect between the peak of elemental Er in Fig. 2 and the Er 4$d$ spectrum at C: it clearly turns out that the satellite peak to main peak intensity ratio is higher. This might hypothetically indicate that Er is not yet entirely metallic and could be very weakly oxidized. Just below C, at marker D, Er is now plainly metallic, as exhibited by the Er $4d$ spectrum in Fig. 3b. Between markers B and D, the occurrence of a slight pile-up of oxygen is observed (see Fig. 1). This contamination could be either attributed to the residual oxidation of the Ti target in the crucible when Ti is evaporated onto the existing Er layer or to the exposition of Er to the residual vacuum in the deposition chamber before Ti deposition. Nevertheless, even with a Ti capping layer as thin as 10 nm, the oxidation of the Er film is very superficial.

---

[2] To be precise, what we call the Er 4$d_{5/2}$ BE throughout the text is that of the main Er 4$d_{5/2}$ peak.

More in depth, at a position referring to marker E, Si shows up, with a corresponding Si $2p$ peak of 98.3 eV, as shown in Fig. 3a. Upon further sputtering, the Si $2p$ BE shows a consequential raise to 98.9 eV at label F, and then increases slowly to reach a plateau of 99.1 eV at marker G. Lollman *et al.* observed a similar shift for the deposition of thin Er layers over Si(111). During the room-temperature deposition of Er, Er and Si atoms can intermix since the Er atoms react strongly with Si (solid-state amorphization[40]). With the increase of the deposited thickness, Si atoms are more diluted into the Er film and are consequently surrounded by a greater number of Er atoms, resulting in a larger Si $2p$ BE shift. Similarly, Wetzel *et al.* reported a variation of the Si $2p$ BE for amorphous mixtures of Er and Si ($Er_xSi_{1-x}$, $x\leq1$), depending on the concentration of Si in Er (large shift of 1.3 eV for $x\approx0.8$). In consequence, the layer between E and G can be very likely identified with an Er/Si mixture of variable composition with an Er/Si intensity ratio that progressively decreases when approaching marker G, as indicated by the concomitant increase of the Si $2p$ BE (see Fig. 3a). However, it is important to draw here the attention on one detail. As already mentioned before, the Si $2p$ BE in the Er/Si film (between E and G) in Fig. 3a exhibits two distinct behaviors: an abrupt variation (E-F) followed by a slow increase (F-G). Indeed, XPS does not only probe the very surface of the sample but also a small thickness below, depending widely on the material under investigation. This effect must all the more be taken into consideration when thin layers are considered. The previous comment is very general and is still valid for the remainder of our analysis. Consequently, we think that the change of behavior in the F-G layer is related to the Si substrate detection. A connection can be made between the Er/Si alloy and the thin amorphous Er/Si layer (a-Er/Si) of 3 nm observed in the high-resolution TEM (HRTEM) picture (inset to Fig. 1). On the contrary, the Er $4d_{5/2}$ BE is constant (167.3 eV) throughout the intermixed layer (see Fig. 3a). A typical Er $4d$ spectrum (extracted at marker F) that we attribute to silicided Er is shown in Fig. 3b with a shift of 0.2 eV towards higher BE after position E. The shapes of the elemental and silicided Er spectra are very similar. Kennou *et al.* even observed no BE difference at all. A weak or negligible shift denotes a weak charge transfer between Si and Er upon alloying, as generally observed for transition metals.

After G, the Si 2$p$ BE of 99.2 eV corresponds to elemental Si, meaning that we reach the Si bulk substrate while the Er 4$d_{5/2}$ typical doublet structure progressively disappears but is still detected (see Fig. 3a). Er is thought to redeposit and even bury into the substrate during the sputtering process due to its heavy mass (knock-on effect[41]).

**B. Ti/Er/Si(100) stack annealed at 300 °C (sample 2)**

From Fig. 4, depicting the XPS intensity depth profile for sample 2 (with markers A' to H'), we can see that, after annealing at 300 °C, a large amount of oxygen diffuses through the whole stack. As previously, the surface (marker A') is composed of TiO$_2$. At label B' (same depth as B), the Er 4$d_{5/2}$ core level is centered at 169.4 eV (instead of 167.7 eV for the as-deposited sample), a BE which is shifted by more than 2 eV when compared to its value for elemental Er. This observation demonstrates that Er is now present in its oxidized form. Concomitantly, the O 1$s$ peak position is shifted from 530.5 eV at position A' to 531 eV at position B', indicating that oxygen is now preferably coupled to Er rather than to Ti. Surprisingly, the Ti 2$p_{3/2}$ BE energy is found to be equal to 454 eV, which can be ascribed to either elemental or slightly oxidized Ti. Those previous results could indicate that, at 300 °C, Er oxidizes much faster and preferentially compared with Ti. As a significant part of oxygen at position B' originates from diffusion through the Ti layer, it is speculated that Er could even reduce Ti oxide: a similar situation was reported by Kennou *et al.* who observed the reduction of SiO$_2$ by Er at 750 °C. Moreover, the preferential formation of Er sesquioxide relatively to Ti dioxide is confirmed by the respective heats of formation of -1897.9 kJmol$^{-1}$ for Er$_2$O$_3$ and -944 kJmol$^{-1}$ for TiO$_2$.[42]

Below position B', we can see that a considerable amount of oxygen penetrates into the cap (Fig. 4), causing a deep oxidation of the Er layer. This assertion is confirmed by the analysis of the Er 4$d_{5/2}$ and O 1$s$ peaks' positions at marker C' (slightly below the cap): BEs of 169.5 and 531.1 eV are recorded, respectively. Between B' and C', Er is completely oxidized and we do not detect the presence of elemental Er anymore. The previous observations constitute evidence that Er is oxidized in depth due to the activation of the diffusion of residual oxygen in the RTA chamber under the application of the thermal budget.

Si shows up at position D' with a weak Si 2$s$ peak centered around 149.5 eV, as we can see in Fig. 5a showing the evolution of the Si 2$s$ and Er 4$d_{5/2}$ BEs versus the sputter time. In comparison with BEs for pure Si (150.5 eV), this value of 149.5 eV suggests the formation of a small Er silicide quantity in oxidized Er or non stoichiometric silicate with a small Si content (ErSi$_x$O$_y$ with x << 1 and y close to 1.5) even though the Er 4$d_{5/2}$ core level at D' firmly remains in the accepted range for oxidized Er (169.5 eV). Nevertheless, when compared to the BE at marker C' (oxidized Er without Si), it appears that the shape of the Er $4d_{5/2}$ peak somewhat transforms (see Fig. 5b), leading to think that a small Er amount could also be alloyed to Si.

Upon further sputtering (positions E' and F'), the characteristic shape of the Er $4d_{5/2}$ peak continues to modify comparatively to marker D' (see Fig. 5b), while the Si $2s$ BE remains roughly unchanged (Fig. 5a). This suggests that, below marker D', the original metallic Er film is converted into a mixture of silicided and oxidized Er or into a silicate, because of the antagonist diffusion of, on the one side, oxygen through the Ti cap and, on the other side, Si in Er in the opposite direction, Si being the diffusive species in reactions with RE elements.[43] The evolution of both the Er $4d_{5/2}$ peak shape and BE probably reflects a progressive change in chemical environment, more specifically, the increase with depth of the Er silicide proportion in the mixture or in the silicate. The O $1s$ BE (531.1 eV) matches the value found for oxidized Er in this work (531 eV) and is also compatible with values for Er silicates (531.1 eV). In consequence, from this analysis, we cannot deduce if we have a mixture of Er silicide and oxidized Er with phase separation or a silicate. But we can affirm that Er, Si, and O are simultaneously present and that the relative (Er-O)/(Er-Si) proportion decreases with depth.

Indeed, at position G' (Fig. 5b), the Er $4d_{5/2}$ peak is now split into a doublet structure very much alike the one depicted in Fig. 3b at marker F, meaning that Er is no more present in its oxidized form and thus in all likelihood silicided. It is worth noting that the doublet already shows up before, at position F', but the global shape of the Er $4d_{5/2}$ core level at that label remains significantly different from that of Er silicide. From a quantitative standpoint, the Er $4d_{5/2}$ BE is recorded at 167.3 eV, BE which is assumed to be related to Er silicide formation as already noted at the interface with Si (marker F) of the previous as-deposited sample. The Si $2s$ BE of 149.8 eV reinforces the assertion that Er appears to be preferably coupled to Si at label G'. Despite the occurrence of some oxygen, there is no contradiction with the previous statement that Er forms a silicide at this position since, as already mentioned, the signature of both materials can be often detected on both sides of an interface.

In Fig. 5a, we can see that below G', as previously observed for sample 1, the Er $4d_{5/2}$ BE is nearly constant (167.3 eV) while the Si $2s$ peak rapidly shifts from 149.8 eV to 150.3 eV (position H') and then slowly reaches 150.5 eV at label I'. Once again, the layer between G' and I' can be assimilated to an Er/Si mixture of variable composition and can be associated to the a-Er/Si layer depicted in the HRTEM inset to Fig. 4. The same parallel can be made between layers H'-I' and F-G: the slow-down of the Si $2p$ BE shift can be linked to the onset of the Si substrate probing. But the comparative inspection of Figs. 1 and 4 suggests that the Er/Si layer is much broader at 300 °C (thickness E-G smaller than thickness F'-I'), as also confirmed by the HRTEM analyses (3 nm versus 6 nm).

**C. Ti/Er/Si(100) stack annealed at 600 °C**

To form a crystalline and stable Er disilicide (ErSi$_{2-x}$) film with the present process conditions, an annealing temperature higher than 300 °C is required, at least 400 °C. To be able better evidencing the potential oxygen diffusion, we set the annealing temperature at 600 °C. Since the stack was already strongly oxidized at 300 °C for a 10 nm thick capping layer and since it was previously shown that a 10 nm thick Ti cap does not provide a good barrier against oxidation at 600 °C, we investigate the increase of the Ti layer thickness (15 nm for sample 3 and 50 nm for sample 4) as a means to improve the protection against oxygen diffusion. Another element that deserves further exploration is the potential impact of the residual oxygen concentration in the RTA chamber during the annealing. To that purpose, an additional sample with a fixed Ti thickness of 10 nm is annealed for a considerably longer pre-RTA purge time of 10 min (sample 5). It is worth recalling here that samples 3 and 4 are grown after a short pre-RTA purge step of 2 min, like samples 1 and 2.

*1. 15 nm thick Ti cap (sample 3)*

As featured in Fig. 6, a lot of oxygen penetrates into the stack and contaminates the top of the Er profile. Without surprise, the surface of the capping layer is composed of TiO$_2$ (marker A''). More in depth, the cap is essentially a mixture of TiO, TiO$_2$, and Ti, with a progressively decreasing TiO$_2$ amount.

A detectable Er signal can be registered from marker B''. In the top of the Er profile (between positions B'' and D''), Er is essentially oxidized (with characteristic O 1$s$ and Er 4$d_{5/2}$ BEs of 531.1 and 169.1 eV, respectively). Throughout the whole B''-D'' region, the Si 2$p$ spectra are characterized by two main peaks with roughly constant BEs at ~103 eV and ~98.8 eV (see Fig. 7a). The first BE testifies to the persistence of stoechiometric Si oxide or Er silicate, both of them possessing similar Si 2$p$ BEs (see Table II). If we have a closer look at the spectrum of the intermediate position C'' in Fig. 7a, we can also notice the presence of several small peaks between the two main ones, probably from Si suboxides or silicates with a lower oxygen content. Comparing the Si 2$p$ spectra at markers B'', C'', and D'' in Fig. 7a, it is also found out that the higher BE peak progressively fades out.

As we know that crystalline ErSi$_{2-x}$ grows at 600 °C [confirmed by HRTEM and XRD analyses (insets to Fig. 6)], we believe that the small but visible ~0.4 eV shift of the second Si 2$p$ peak relatively to the elemental Si peak can be attributed to ErSi$_{2-x}$ (see Fig. 7a). In any case, it is very different from the Si 2$p$ Er silicate peak given in the literature (see Table II). Between B'' and D'', the ErSi$_{2-x}$ Si 2$p$ peak becomes steadily predominant comparatively to the other peaks, reflecting a progressive increase of the ErSi$_{2-x}$ proportion (see Fig. 7a). Before label D'', the Er 4$d_{5/2}$ core level of ~169.1 eV can be unambiguously attributed to oxidized Er. Still, its shape shows a drastic modification between C'' and D'' (Fig. 7b), comparable to the one observed for sample 2. We can conclude that the layer between B'' and D'' presents XPS signatures of ErSi$_{2-x}$ and probably Er silicate, as further confirmed by the weak peak detected by XRD (inset to Fig. 6). The layer could be seen as oxidized ErSi$_{2-x}$ with a variable composition, containing less and less oxygen as and when one penetrates more deeply into the stack. The O 1$s$ BE of 531 eV is in agreement with Er silicate BEs from the literature. It is worth mentioning here that we cannot give accurate BEs for a well-identified Er silicate since it has no fixed composition. Even though it not easy to determine the exact composition of the film between B'' and D'', it is clear that it is not composed of pure ErSi$_{2-x}$ and that oxygen substantially contaminates the top of the ErSi$_{2-x}$ layer.

The typical doublet structure of Er silicide appears at marker D'' and the corresponding Er 4$d_{5/2}$ core level drops to 167.5 eV (i.e. the main Er 4$d_{5/2}$ peak becomes dominant). It is interesting at this point to make a comparison with amorphous Er silicide. In both Er silicide types, the Er 4$d_{5/2}$ signals are very similar and the Er 4$d_{5/2}$ BE is fixed to 167.3 eV (Figs. 3a and 5a versus Fig. 7c). On the other side, the Er/Si intensity ratio is approximately constant for ErSi$_{2-x}$, contrary to a-Er/Si (Figs. 1 and 4 versus Fig. 6). Another substantial difference is the evolution of the Si 2$p$ core level before the detection of the Si substrate (between D'' and E''): it is constant (98.7 eV) for ErSi$_{2-x}$ while it rises rapidly for a-Er/Si (E-F versus F'-H'). This difference can be simply explained by the fact that ErSi$_{2-x}$ has a fixed stoichiometry whereas a-Er/Si is characterized by a variable composition.

Finally, the Si substrate is reached in F''. It is worth noting that a strong Er signal is still observed at this position due to the knock-on of Er atoms into the Si substrate, as already observed for the two other samples. Once again, at the vicinity with the interface to Si, the silicidation reaction is predominant (D''-F'') even if a large oxygen amount diffuses through the Ti cap and oxidizes the top of the ErSi$_{2-x}$ layer (B''-D''). The capability of growing oxygen-free ErSi$_{2-x}$ in the vicinity of the Si interface is further confirmed by the extraction of a low SBH of 0.295 eV. More details about the extraction procedure can be found in Ref. 8 and Ref. 44. The corresponding experimental Arrhenius plot (blue circle markers) and its fit (blue solid lines) are shown in Fig. 8.

Moreover, it is observed that the fit is excellent over the whole temperature range, supplementary testimony of the contact quality. The current-voltage measurements are performed between 150 and 290 K with a step of 20 K, for biases ranging from 0.1 V up to 1 V with a step of 0.15 V. This result illustrates the robustness of low barrier contact formation over *n*-Si based on ErSi$_{2-x}$ since a SBH very close to the state-of-the-art value is obtained in spite of the nonideal manufacturing conditions (high vacuum deposition and residual oxygen in the annealing atmosphere).

*2. 50 nm thick Ti cap (sample 4)*

In Fig. 9a, we show the intensity depth profile for the last part of the capping layer only and the ErSi$_{2-x}$ film. The part of the cap in immediate contact with the ErSi$_{2-x}$ film is composed of metallic Ti (from A*) while the top of the cap is as before a mixture of Ti dioxide and suboxides. TEM micrographs in Fig. 9b and c illustrate the drastic transformation of the Ti cap subjected to RTA at 600 °C. Before annealing (Fig. 9b), the Ti layer is 50 nm thick and covered by a thin TiO$_2$ layer (as pointed out for sample 1). After heating, the cap gets considerably thicker (~90 nm) and is divided into two distinct layers: on the top, TiO$_2$, and very weakly oxidized Ti at the bottom.

The important result to pinpoint is that we observe no formation of oxidized ErSi$_{2-x}$ due to oxygen diffusion through the Ti cap (Er 4$d_{5/2}$ BE = 167.3 eV throughout the ErSi$_{2-x}$ film). This is also confirmed by the examination of the O 1*s* signal which is extinguished after label B*. This proves that growth under vacuum is not mandatory to produce oxygen-free RE silicides. Still, there is a slight oxygen pile-up between Ti and ErSi$_{2-x}$ (around C*). Most probably, it does not originate from oxygen diffusion but is rather a trace of the pile-up as noted for sample 1. The characteristic doublet of ErSi$_{2-x}$ is observed starting from position C*, is visible and well-defined up to position E*. Over the interval between markers C* and D*, the Si 2*s* BE is centered on ~149.9 eV. Beyond position D*, the Si 2*s* BE begins to shift towards the BE of elemental Si. The Si substrate is thus probed from that position and is reached in position E* where the Si 2*s* BE matches that of elemental Si (150.5 eV). ErSi$_{2-x}$ is thus present between C* and E*. For comparison with sample 3, the Arrhenius plot of sample 4 is also given in Fig. 8 (in red) and the corresponding SBH amounts to 0.3 eV. It is worth pointing out that, even if sample 3 is contaminated, its SBH is very close to that of sample 4, additional proof of the preserved silicide/Si interface. It is worth making here a short comment about the formation of structural defects, like pyramids,[45] that are well known to deteriorate the morphology of RE silicide films. Indeed, for samples 3 and 4, we notice the occurrence of pyramidal defects, regardless of the initial Ti capping thickness. This means that the Ti capping layer does not hinder their formation. Still, the pyramids demonstrate no apparent effect on the electrical characteristics of the ErSi$_{2-x}$/$n$-Si contacts, notably on the SBH homogeneity.

*3. 10 nm thick Ti cap with a 10 min long purge (sample 5)*

Comparing intensity profiles of Figs. 6 and 10, it is clearly observed that the Ti oxidation is considerably reduced for a significantly longer purge of 10 min (even for a slightly thinner cap). The corresponding Ti 2$p_{3/2}$ BEs are typical of slightly oxidized Ti at the top of the capping layer (454.7 eV) and of elemental Ti deeper into the layer (453.9 eV). In addition, the cap exhibits no volumetric expansion, contrary to samples 3 and 4: the cap thickness is proportional to a sputtering time of 50 min before and after annealing (comparing Figs. 1 and 10) while the corresponding sputter time is 150 min for sample 3 (Fig. 6). Despite the weak oxidation of the cap, the top of the ErSi$_{2-x}$ layer is still oxidized. As already suggested here above, it is plausible to think that, during silicidation, a small amount of residual oxygen diffuses into the Ti cap and that Er pumps that oxygen away from the cap, thereby reducing the Ti layer. BEs related to pure ErSi$_{2-x}$ (167.4 eV) are recorded from position A**. In the ErSi$_{2-x}$ layer, the Si 2$s$ BE is roughly constant (150±0.1 eV) and is characteristic of ErSi$_{2-x}$. From position B**, the Si 2$s$ BE starts to shift, meaning that the Si substrate is close. The substrate is reached at marker C** and the interface is located around that position.

*4. 50 nm thick Ti cap with a 10 min long purge (sample 6)*

In the context of SBMOSFET fabrication, the stripping of the capping layer must be further addressed. An efficient recipe to etch Ti compounds is based on a mixture of NH$_4$OH, H$_2$O$_2$ and H$_2$O. The recipe is first experimented on samples 3 and 4. The capping layer proves particularly hard to remove, probably because it is strongly oxidized. In consequence, sample 6 is prepared in the optimal conditions as determined here above, to obtain oxygen-free ErSi$_{2-x}$ and weakly oxidized Ti. Figure 11 exhibits an XRD spectrum and, in inset, a TEM cross-section of sample 6, after stripping with the same solution. Both analyses testify to the preservation of the ErSi$_{2-x}$ film that reveals a smooth morphology. Consequently, a supplementary advantage of the long purge is to reduce the cap oxidation and thereby to facilitate its stripping. The extracted SBH amounts to 0.303 eV, manifesting no degradation after selective removal.

## Conclusions

In the present work, the chemical state of various Ti/Er/Si(100) stacks subjected to different RTA conditions is examined in detail by means of XPS, with a peculiar focus on the chemical changes of Er. The influence of the annealing temperature, the Ti cap thickness and the pre-annealing purge duration is investigated. In the case of the as-deposited stack, the Er film is found to be oxygen-free, except for some superficial oxidation. A shift of the Si spectra close to the interface with Si is disclosed, suggesting the onset of silicidation. At 300 °C, the Er film is nearly completely oxidized, except in the vicinity of the interface where silicidation appears predominant. For both cases, the formation of amorphous Er silicide is confirmed by HRTEM. At 600 °C, the formation of $ErSi_{2-x}$ is clearly identified by XRD and HRTEM. A significant part of the $ErSi_{2-x}$ film is revealed to contain oxidized Er if the pre-annealing purge is too short. However, in the surroundings of the interface to Si, XPS shows that $ErSi_{2-x}$ seems preserved, as also testified by a low SBH of 0.3 eV. An improvement is observed for a long purge but the top of the $ErSi_{2-x}$ layer is still contaminated. Finally, provided a thicker 50 nm Ti film covers the Er layer, an oxygen-free $ErSi_{2-x}$ film is grown and the cap stripping facilitated if the pre-RTA purge is long.

From all the previous results, we can summarize that the Ti cap provides sufficient protection to allow for silicidation, which always appears predominant over oxidation near the Si interface, regardless of the annealing conditions.

## Acknowledgements

This work was supported by the European Commission through the NANOSIL network of excellence (FP7-IST-NoE-216171) and the METAMOS project (FP6-IST-STREP-016677).

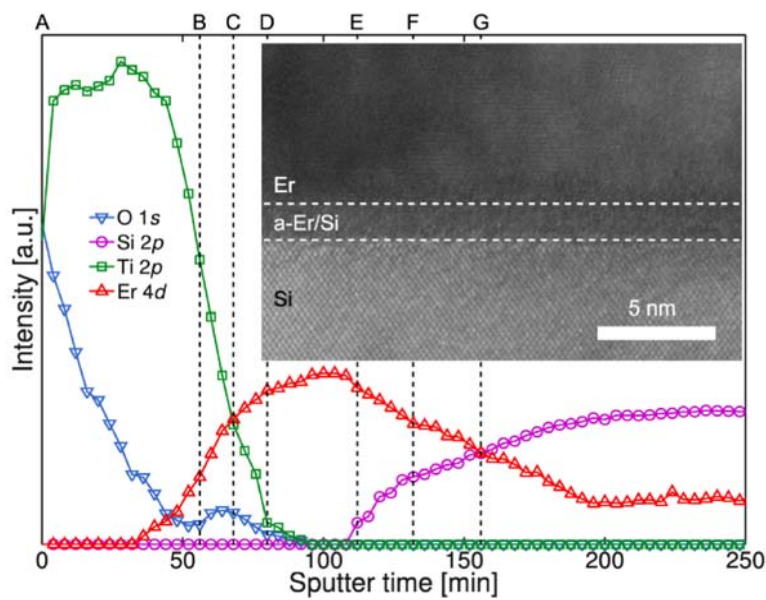

**Figure 1.** (Color online) XPS intensity depth profile of the as-deposited Ti(10 nm)/Er(25 nm)/Si(100) stack. Particular positions along the profile are marked by labels A to G. Inset: Cross-sectional HRTEM picture of the interface between a-Er/Si and Si.

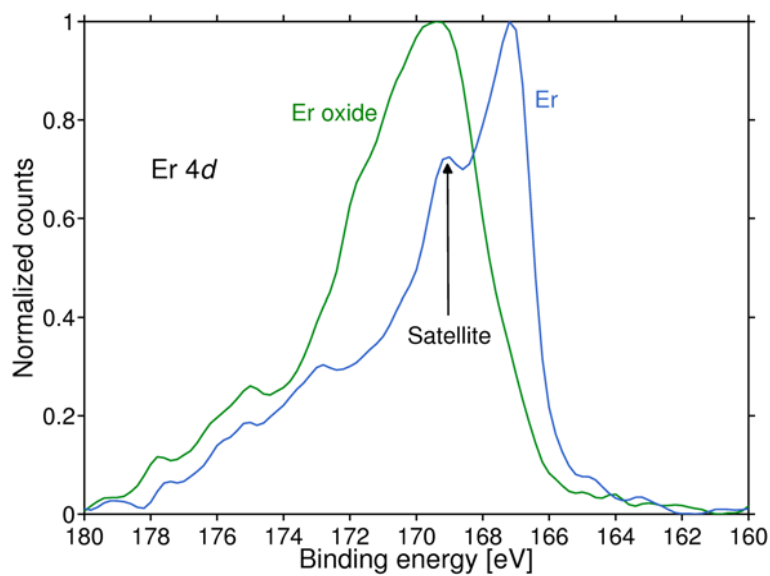

**Figure 2.** (Color online) Typical normalized Er $4d$ core level spectra after background subtraction for metallic and oxidized Er, respectively. The spectra are obtained from a thick Er layer to serve as a reference. The Er $4d_{5/2}$ main peak of elemental Er is located at 167.1 eV while its satellite is at 169.2 eV. Oxidized Er is characterized by a unique peak at 169.3 eV.

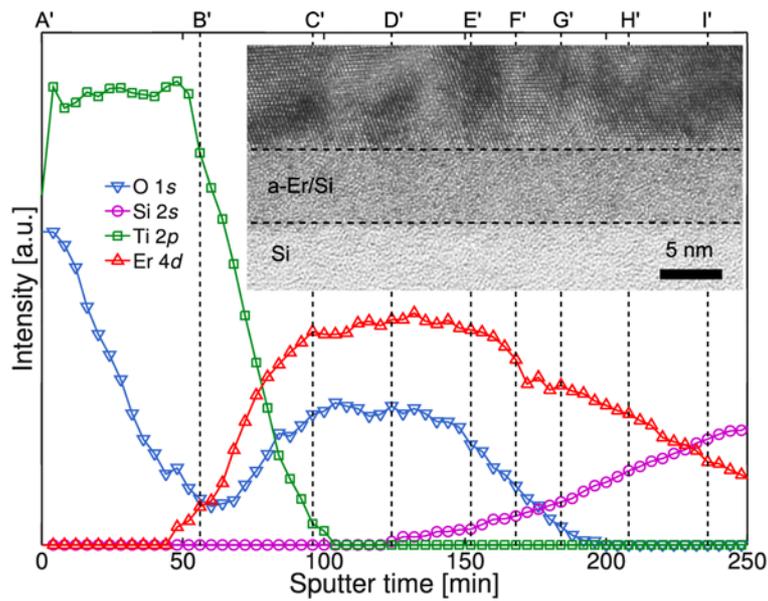

**Figure 3.** (Color online) (a) Er $4d_{5/2}$ and Si $2p$ BEs for the as-deposited Ti(10 nm)/Er(25 nm)/Si(100) stack, from marker C. (b) Normalized Er $4d$ core level spectra after background subtraction for the same stack, for markers C to F.

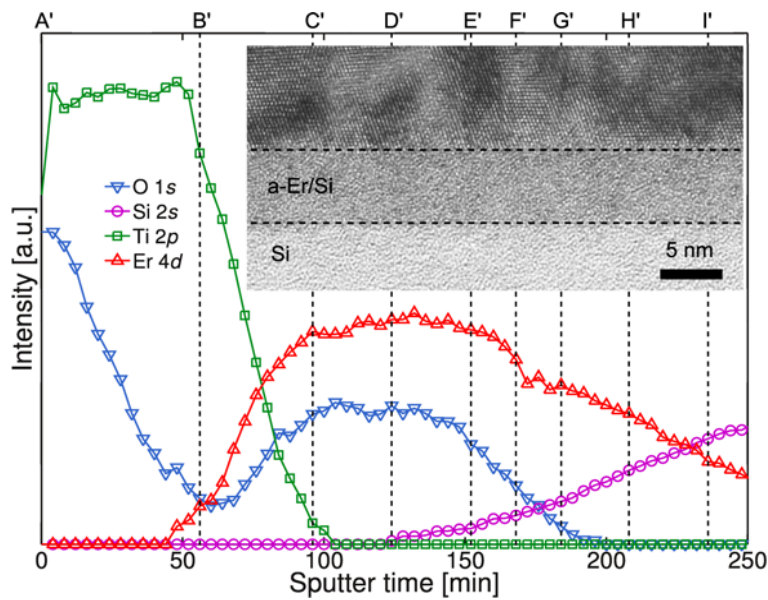

**Figure 4.** (Color online) XPS intensity depth profile of the Ti(10 nm)/Er(25 nm)/Si(100) stack annealed at 300 °C for 2 min with a 2 min pre-RTA purge. Particular positions along the profile are marked by labels A' to I'. Inset: Cross-sectional HRTEM picture of the interface between a-Er/Si and Si.

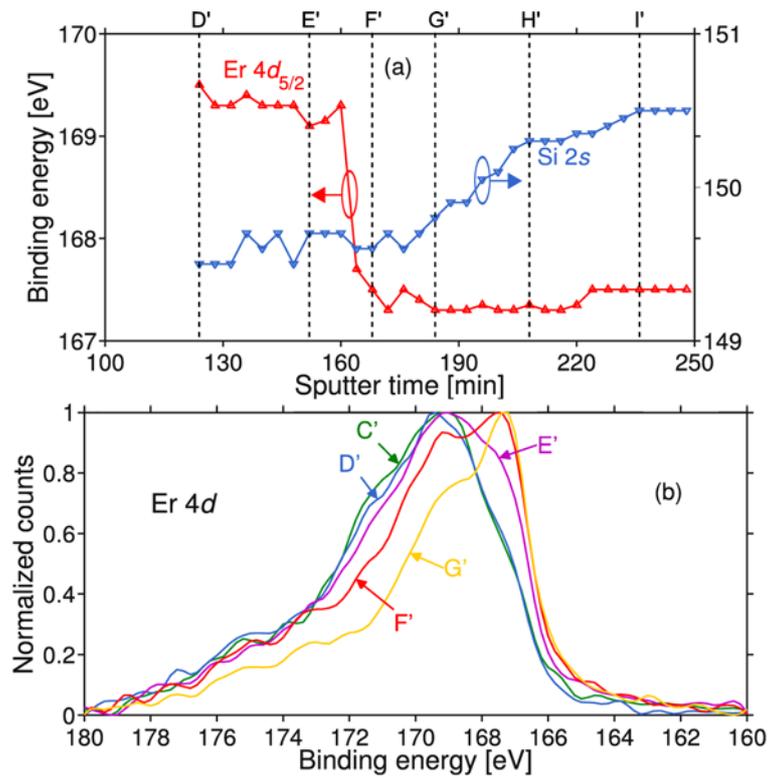

**Figure 5.** (Color online) (a) Er $4d_{5/2}$ and Si $2s$ BEs for the Ti(10 nm)/Er(25 nm)/Si(100) stack annealed at 300 °C for 2 min with a 2 min pre-RTA purge, from marker D'. (b) Normalized Er $4d$ core level spectra after background subtraction for the same stack, for markers C' to G'.

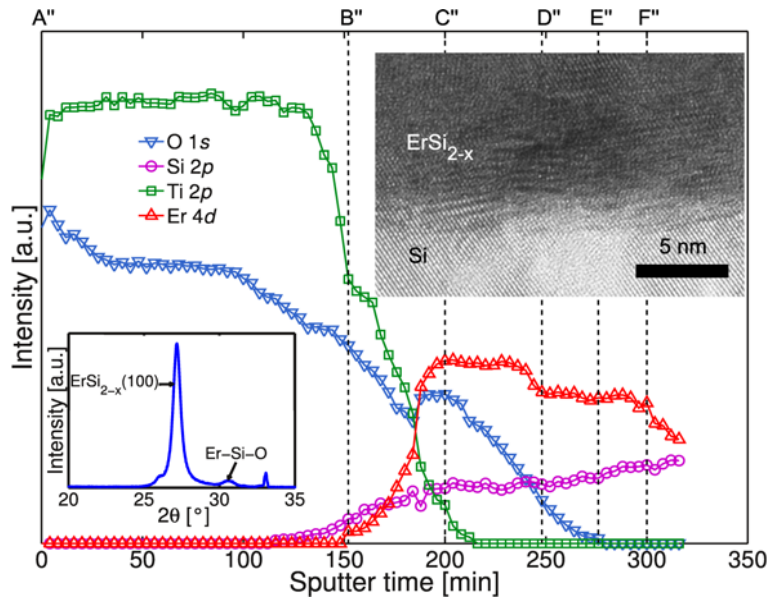

**Figure 6.** (Color online) XPS intensity depth profile of the Ti(15 nm)/Er(25 nm)/Si(100) stack annealed at 600 °C for 2 min with a 2 min pre-RTA purge. Particular positions along the profile are marked by labels A'' to F''. Inset 1: Cross-sectional HRTEM picture of the interface between $ErSi_{2-x}$ and Si. Inset 2: XRD spectrum with a strong peak corresponding to $ErSi_{2-x}$.

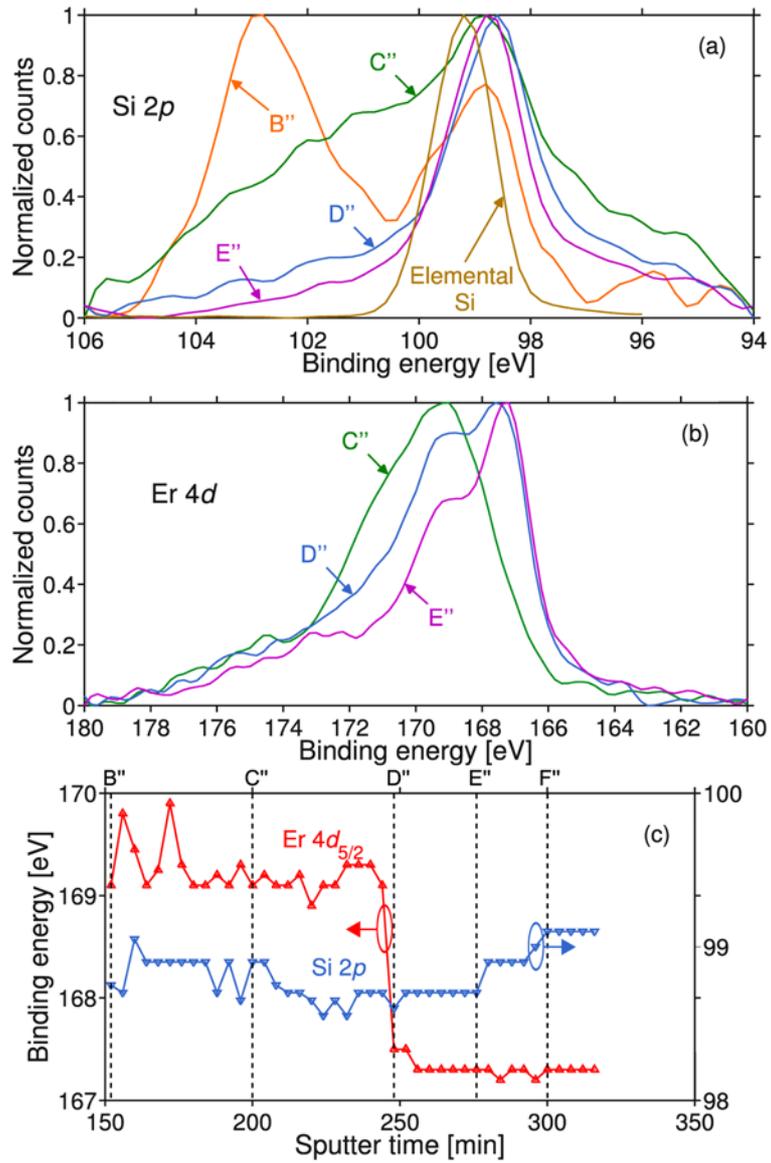

**Figure 7.** (Color online) Normalized (a) Si 2*p* and (b) Er 4*d* core level spectra after background subtraction for the Ti(15 nm)/Er(25 nm)/Si(100) stack annealed at 600 °C for 2 min with a 2 min pre-RTA purge. (c) Er $4d_{5/2}$ and Si 2*p* BEs for the same stack, from marker B''.

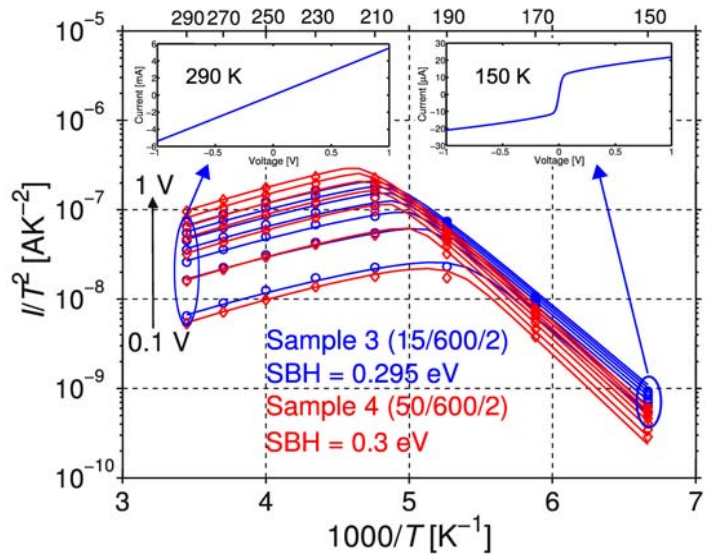

**Figure 8.** (Color online) Experimental (markers) and modeled (solid lines) Arrhenius plots for the Ti(15 nm)/Er(25 nm)/Si(100) stack annealed at 600 °C for 2 min with a 2 min pre-RTA purge (blue) and for the Ti(50 nm)/Er(25 nm)/Si(100) stack annealed at 600 °C for 2 min with a 2 min pre-RTA purge (red). The current-voltage measurements are performed between 150 and 290 K with a step of 20 K, for biases ranging from 0.1 V up to 1 V with a step of 0.15 V. The average error on the extracted SBH lies within 5 meV. Insets: Typical experimental current-voltage characteristics at 150 and 290 K.

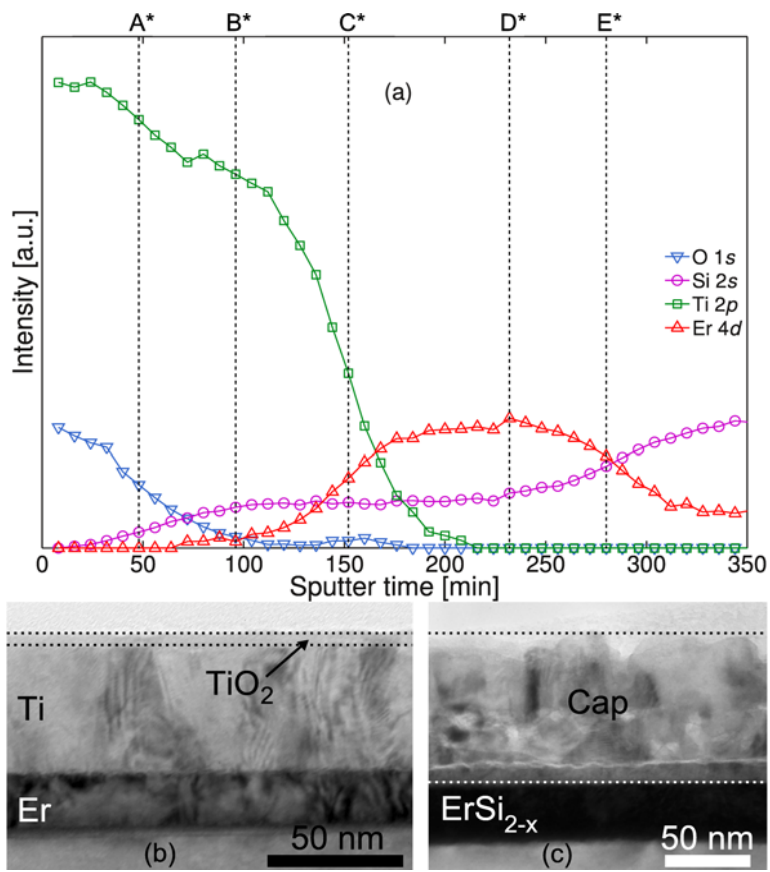

**Figure 9.** (Color online) (a) XPS intensity depth profile of the Ti(50 nm)/Er(25 nm)/Si(100) stack annealed at 600 °C for 2 min with a 2 min pre-RTA purge. Particular positions along the profile are marked by labels A* to E*. TEM micrographs of the same stack (b) left as-deposited and (c) after annealing.

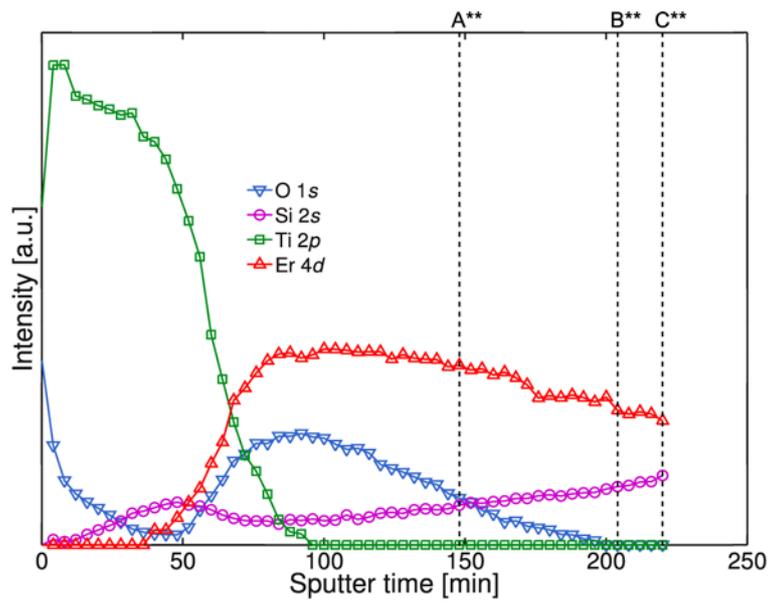

**Figure 10.** (Color online) XPS intensity depth profile of the Ti(10 nm)/Er(25 nm)/Si(100) stack annealed at 600 °C for 2 min with a 10 min pre-RTA purge. Particular positions along the profile are marked by labels A** to C**.

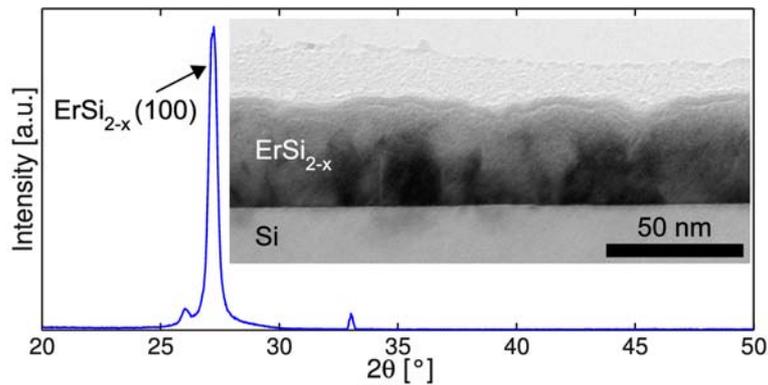

**Figure 11.** (Color online) XRD spectrum of the Ti(50 nm)/Er(25 nm)/Si(100) stack annealed at 600 °C for 2 min with a 10 min pre-RTA purge. Inset: Corresponding cross-sectional TEM picture.

Table I: Summary of the stack composition and RTA conditions for all considered samples.

| Sample | Ti thickness [nm] | Annealing temperature [°C] | Pre-RTA purge duration [min] |
|---|---|---|---|
| 1 (10/no/2) | 10 | no | 2 |
| 2 (10/300/2) | 10 | 300 | 2 |
| 3 (15/600/2) | 15 | 600 | 2 |
| 4 (50/600/2) | 50 | 600 | 2 |
| 5 (10/600/10) | 10 | 600 | 10 |
| 6 (50/600/10) | 50 | 600 | 10 |

Table II: XPS BEs relevant to Er $4d_{5/2}$, O $1s$, Si $2p/2s$, and Ti $2p_{3/2}$ core levels stemming from the literature and determined in this work.

| Peak | Compound | BE [eV] | Reference |
|---|---|---|---|
| Er $4d_{5/2}$ | Er | 167.4+169 (sat.) | Swami |
| | | 167.6 | Kennou |
| | | 167.1 | Reckinger |
| | | **167.1** | This work |
| | ErSi$_{2-x}$ | 167.6 | Kennou |
| | | **167.3** | This work |
| | a-Er/Si | 167.3 | Reckinger |
| | | **167.3** | This work |
| | Er$_2$O$_3$ | 168.7 | Uwamino |
| | | 168.6 | Swami |
| | | 168.8 | Wagner |
| | | 170.4 | Guerfi, Kennou |
| | ErO$_x$ | 169.2 | Kennou |
| | | 169.5 | Reckinger |
| | | **169.1-169.5** | This work |
| | ErSi$_x$O$_y$ | 170 | Hafidi |
| | | 169-169.7 | Choi |
| O $1s$ | Er$_2$O$_3$ | 530.6 | Swami |
| | | 531.6 | Guerfi, Netzer |
| | ErO$_x$ | **531** | This work |
| | SiO$_x$ (x ≈ 1) | 532-533 | Kennou |
| | SiO$_2$ | 533.2 | Kennou |
| | | 533.1 | Larrieu |
| | TiO$_2$ | 529.3-531 | Atuchin |
| | | 530.5 | Demri |
| | | 529.8 | Moses |
| | ErSi$_x$O$_y$ | 531.1 | Pan |
| Si $2s$ | Si | 150.7 | Ogama |
| | | 150.5-150.6 | Reckinger |
| | | **150.5** | This work |
| | SiO$_2$ | 154.2 | Anpo |
| | ErSi$_{2-x}$ | **150.1** | This work |
| | a-Er/Si | 149.5 | Reckinger |
| | | **149.6** | This work |
| Si $2p$ | Si | 99.3 | Lollman, Gokhale |
| | | **99.1** | This work |
| | ErSi$_{2-x}$ | 99 | Lollman |
| | | **98.7** | This work |
| | a-Er/Si | **98.1** | Lollman |
| | | Shift: 0 to 1.3 | Wetzel |
| | | 98.3 | This work |
| | SiO$_2$ | ~103 | Guerfi |
| | | ~103.3 | Hafidi |
| | SiO$_x$ | ~102 | Guerfi |
| | ErSi$_x$O$_y$ | ~103.8 | Hafidi |
| Ti $2p_{3/2}$ | Ti | 454 | Atuchin |
| | | 454 | Badrinarayanan |
| | TiO$_2$ | 459 | Atuchin |
| | | 458.9 | Gonbeau |
| | | 458.9 | Demri |

Table III: Summary of the Ti $2p_{3/2}$, O $1s$, Er $4d_{5/2}$, and Si $2p/2s$ BEs [eV] recorded at the different markers and corresponding composition, for samples 1 to 5.

| Sample | Marker | Sputter time [min] | Ti $2p_{3/2}$ | O $1s$ | Er $4d_{5/2}$ | Si $2p/2s$ | Composition |
|---|---|---|---|---|---|---|---|
| 1 (10/no/2) | A | 0 | 458.9 | 530.5 | X | X | $TiO_2$ |
| | B | 56 | 453.7 | 530.9 | 167.7 | X | Ti + Er |
| | C | 68 | 453.7 | 530.7 | 167.1 | X | $ErO_x$ |
| | D | 80 | 453.5 | 530.7 | 167.1 | X | Er |
| | E | 112 | X | X | 167.1 | 98.3 | a-Er/Si |
| | F | 132 | X | X | 167.3 | 98.9 | a-Er/Si |
| | G | 156 | X | X | 167.3 | 99.1 | Si |
| 2 (10/300/2) | A' | 0 | 459 | 530.5 | X | X | $TiO_2$ |
| | B' | 56 | 454 | 530.9 | 169.4 | X | Ti + $ErO_x$ |
| | C' | 96 | 454 | 531.1 | 169.5 | X | $ErO_x$ |
| | D' | 124 | X | 531.1 | 169.5 | 149.5 | Er-Si-O |
| | E' | 152 | X | 531.1 | 169.1 | 149.7 | Er-Si-O |
| | F' | 168 | X | 531 | 167.5 | 149.6 | Er-Si-O |
| | G' | 184 | X | 530.7 | 167.3 | 149.8 | a-Er-Si |
| | H' | 208 | X | X | 167.3 | 150.3 | a-Er-Si |
| | I' | 236 | X | X | 167.5 | 150.5 | Si |
| 3 (15/600/2) | A'' | 0 | 459.1 | 530.5 | X | X | $TiO_2$ |
| | B'' | 152 | 455.1 | 531.1 | 169.1 | 102.9/98.8 | Er-Si-O |
| | C'' | 200 | 454.1 | 531 | 169.1 | 98.9 | Er-Si-O |
| | D'' | 248 | X | 530.9 | 167.5 | 98.6 | Er-Si-O |
| | E'' | 276 | X | 530.3 | 167.3 | 98.7 | $ErSi_{2-x}$ |
| | F'' | 300 | X | X | 167.3 | 99.1 | Si |
| 4 (50/600/2) | A* | 48 | 454 | 531.4 | X | 149.7 | $TiO_x$ |
| | B* | 96 | 453.8 | 531.2 | X | 149.8 | Ti |
| | C* | 152 | 453.8 | 530.9 | 167.3 | 149.9 | Ti + $ErSi_{2-x}$ |
| | D* | 232 | X | X | 167.3 | 150 | $ErSi_{2-x}$ |
| | E* | 280 | X | X | 167.3 | 150.5 | Si |
| 5 (10/600/10) | A** | 148 | X | 531.1 | 167.5 | 150.1 | Er-Si-O |
| | B** | 204 | X | X | 167.3 | 150.2 | $ErSi_{2-x}$ |
| | C** | 220 | X | X | 167.3 | 150.5 | Si |